\input ./style/arxiv-general.cfg
\documentclass[aoas,MSNbibl,nameyear,dvips]{arximspdf}
\makeatletter
   \@ifpackageloaded{graphicx}{}{\usepackage{graphicx}}
\makeatother
\usepackage{dcolumn}

\doi{10.1214/14-AOAS765}
\volume{9}
\issue{2}
\pubyear{2015}
\firstpage{849}
\lastpage{865}
\docsubty{FLA}
\setattribute{copyright}{owner}{\textup{In the Public Domain}}

\makeatletter
\newcolumntype{d}[1]{D{.}{.}{#1}}
\newproclaim{defn}{Definition}[section]
\newproclaim{exam}[defn]{Example}
\newproclaim{rmk}{Remark}
\def\leqslant{\leq}
\def\geqslant{\geq}
\makeatother

\begin{document}
\begin{frontmatter}

\title{A two-state mixed hidden Markov model for risky teenage driving
behavior}
\runtitle{A mixed hidden Markov model for risky driving}

\begin{aug}
\author[A]{\fnms{John C.}~\snm{Jackson}\thanksref{M1}\ead[label=e1]{john.jackson@usma.edu}},
\author[B]{\fnms{Paul S.}~\snm{Albert}\corref{}\thanksref{M2,T1}\ead[label=e2]{albertp@mail.nih.gov}}
\and
\author[B]{\fnms{Zhiwei}~\snm{Zhang}\thanksref{M2,T1}}
\thankstext{T1}{Supported by the
Intramural Research program of the National Institutes of Health,
Eunice Kennedy Shriver National Institute of Child Health and Human
Development. List Paul~S. Albert as corresponding author.
Correspondence should be sent to Paul S. Albert at \printead{e2}.}
\runauthor{J.~C. Jackson, P.~S. Albert and Z. Zhang}
\affiliation{United States Military Academy\thanksmark{M1} and Eunice
Kennedy Shriver National Institute of Child Health and Human
Development\thanksmark{M2}}
\address[A]{J. C. Jackson\\
Department of Mathematical Sciences\\
United States Military Academy\\
Thayer Hall\\
West Point, New York 10996\\
USA\\
\printead{e1}}
\address[B]{P.~S. Albert\\
Z. Zhang\\
Biostatistics and Bioinformatics\\
Division of Intramural Population\\
\quad Health Research\\
Eunice Kennedy Shriver National Institute\\
\quad of Child Health and Human Development\\
6100 Executive Blvd\\
Bethesda, Maryland 20892\\
USA\\
\printead{e2}}
\end{aug}

%
\received{\smonth{2} \syear{2013}}
%
\revised{\smonth{5} \syear{2014}}

%
\begin{abstract}
This paper proposes a joint model for longitudinal binary and count
outcomes. We apply the model to a unique longitudinal study of teen
driving where risky driving behavior and the occurrence of crashes or
near crashes are measured prospectively over the first 18 months of
licensure. Of scientific interest is relating the two processes and
predicting crash and near crash outcomes. We propose a two-state mixed
hidden Markov model whereby the hidden state characterizes the mean for
the joint longitudinal crash/near crash outcomes and elevated g-force
events which are a proxy for risky driving. Heterogeneity is introduced
in both the conditional model for the count outcomes and the hidden
process using a shared random effect. An estimation procedure is
presented using the \textit{forward--backward} algorithm along with
adaptive Gaussian quadrature to perform numerical integration. The
estimation procedure readily yields hidden state probabilities as well
as providing for a broad class of predictors.
\end{abstract}

%
\begin{keyword}
\kwd{Adaptive quadrature}
\kwd{hidden Markov model}
\kwd{joint model}
\kwd{random effects}
\end{keyword}
\end{frontmatter}
%

\section{Introduction}
\label{sec1}

The Naturalistic Teenage Driving Study (NTDS), sponsored by the
National Institute of Child Health and Human Development (NICHD), was
conducted to evaluate the effects of experience on teen driving
performance under various driving conditions. It is the first
naturalistic study of teenage driving and has given numerous insights
to the risky driving behavior of newly licensed teens that includes
evidence that risky driving does not decline with experience as
discussed by \citet{Bruce}. During the study 42 newly licensed drivers
were followed over the first 18 months after obtaining a license. The
participants were paid for their participation, and there were no
dropouts. Driving took place primarily in southern Virginia among small
cities and rural areas. For each trip, various kinematic measures were
captured. A lateral accelerometer recorded driver steering control by
measuring g-forces the automobile experiences. These recordings provide
two kinematic measures: lateral acceleration and lateral deceleration.
A longitudinal accelerometer captured driving behavior along a straight
path and records accelerations or decelerations. Another measure for
steering control is the vehicle's yaw rate, which is the angular
deviation of the vehicle's longitudinal axis from the direction of the
automobile's path. Each of these kinematic measures was recorded as
count data as they crossed specified thresholds that represent normal
driving behavior. Crash and near crash outcomes were recorded in two
ways. First, the driver of each vehicle had the ability to self report
these events. Second, video cameras provided front and rear views from
the car during each trip. Trained technicians analyzed each trip the
driver took using the video and determination of crash/near crash
events made. Table~\ref{summary} shows the aggregate data for the
driving study. More information on the study can be found at
\url{http://www.vtti.vt.edu}. Our interests are
the prediction of crash and near crash events from longitudinal risky
driving behavior. Crash or near crash outcomes are our binary outcome
of interest, while excessive g-force events are our proxy for risky driving.
It is likely that crash/near crash outcomes are best described by some
unobserved or latent quality like inherent driving ability. Previously,
\citet{Jackson} analyzed the driving data using a latent construct
where the previously observed kinematic measures describe the hidden
state and the hidden state describes the CNC outcome. Our approach here
characterizes the joint distribution of crash/near crash and kinematic
outcomes using a mixed hidden Markov model where both outcomes
contribute to the calculation of the hidden state probabilities.

\begin{table}[b]
\tabcolsep=0pt
\caption{Kinematic measures and correlation with CNCs, naturalistic
teenage driving study, $^\ast$correlation computed between the CNC and
elevated g-force event rates}\label{summary}
\begin{tabular*}{\textwidth}{@{\extracolsep{\fill}}lcd{5.0}d{3.1}c@{}}
\hline
\multicolumn{1}{@{}l}{\textbf{Category}} &
\multicolumn{1}{c}{\textbf{Gravitation force}} &
\multicolumn{1}{c}{\textbf{Frequency}}& \multicolumn{1}{c}{$\bolds{\%}$ \textbf{Total events}} &
\multicolumn{1}{c@{}}{\textbf{Correlation with CNCs}$\bolds{^\ast}$}\\
\hline
Rapid starts & $> 0.35$ & 8747 & 39.6 &0.28\\
Hard stops & $\leqslant-0.45$ & 4228 & 19.1 &0.76\\
Hard left turns & $\leqslant-0.05$ & 4563 & 20.6 &0.53\\
Hard right turns& $\geqslant0.05$ & 3185 & 14.4 &0.62\\
Yaw & $6^\circ$ in 3 seconds & 1367 & 6.2 &0.46 \\[3pt]
Total & &\multicolumn{1}{c}{22,090} &\multicolumn{1}{c}{100} &\multicolumn{1}{c}{0.60}\\
\hline
\end{tabular*}
\end{table}

There is a previous literature on mixed hidden Markov models.
Discrete-time mixed Markov latent class models are introduced by \citet
{Landv}. A general framework for implementing random effects in the
hidden Markov model is discussed by \citet{Altman}. In this work, the
author presented a general framework for a mixed hidden Markov model
with a single outcome. The mixed hidden Markov model presented by
Altman unifies existing hidden Markov models for multiple processes,
which provides several advantages. The modeling of multiple processes
simultaneously permits the estimation of population-level effects as
well as allowing great flexibility in modeling the correlation
structure because they relax the assumption that observations are
independent given the hidden states. There are a variety of methods
available for estimation of parameters in mixed hidden Markov models.
\citet{Altman} performed estimation by evaluating the likelihood as a
product of matrices and performing numerical integration via Gaussian
quadrature. A quasi-Newton method is used for maximum likelihood
estimation. \citet{Bart2} extend the latent class model for the
analysis of capture-recapture data, which takes into account the effect
of past capture outcomes on future capture events. Their model allows
for heterogeneity among subjects using multiple classes in the latent
state. \citet{Scott} introduces Bayesian methods for hidden Markov
models which was used as a framework to analyze alcoholism treatment
trial data [\citet{Shirley}]. \citet{Bart1} use a fixed effects model to
evaluate the performance of nursing homes using a hidden Markov model
with time-varying covariates in the hidden process. \citet{Maruotti}
discusses mixed hidden Markov models and their estimation using the
expectation--maximization algorithm and leveraging the \textit{forward} and
\textit{backward} probabilities given by the \textit{forward--backward}
algorithm and partitioning the complete-data-log-likelihood
into sub-problems.
\citet{MaruottiRocci}
proposed
a mixed non-homogeneous hidden
{M}arkov model for a categorical data.

We present a model that extends the work of \citet{Altman} in several
ways. First, we allow the hidden state to jointly model longitudinal
binary and count data, which in the context of the driving study
represent crash/near crash events and kinematic events, respectively. We
also introduce an alternative method to evaluate the likelihood by
first using the \textit{forward--backward} algorithm [\citet{Baum}] followed
by performing integration using adaptive Gaussian quadrature.
Implementation of the \textit{forward--backward} algorithm allows for easy
recovery of the posterior hidden state probabilities, while the use of
adaptive Gaussian quadrature alleviates bias of parameter estimates in
the hidden process [\citet{Altman}]. Understanding the nature of the
hidden state at different time points was of particular interest to us
in our application to teenage driving, and our estimation procedure
yields an efficient way to evaluate the posterior probability of state
occupancy.

In this paper we first introduce a joint model for crash/near crash
outcomes and kinematic events which allows the mean of each these to
change according to a two-state Markov chain. We introduce
heterogeneity in the hidden process as well as the conditional model
for the kinematic events via a shared random effect. We then discuss an
estimation procedure whereby the likelihood is maximized directly and
estimation of the hidden states is readily available by incorporating
the \textit{forward--backward} algorithm [\citet{Baum}] in evaluating
individual likelihoods.
We apply our model to the NTDS data and show that these driving
kinematic data and CNC events are closely tied via the latent state. We
use our model results to form a predictor for future CNC events based
on previously observed kinematic data.

\section{Methods}
\label{sec2}

Here we present the joint model for longitudinal binary and count
outcomes using two hidden states as well as the estimation procedure
for model parameters.
\subsection{The model}
Let $\mathbf{b}_i=(b_{i1},b_{i2},\ldots,b_{in_i})$ be an unobserved
binary random vector whose elements follow a two-state Markov chain
(state ``0'' represents a \textit{good} driving state and state ``1''
represents a \textit{poor} driving state) with unknown transition
probabilities $\mathrm{p}_{01}, \mathrm{p}_{10}$ and initial
probability distribution $r_0$. We model the crash/near crash outcome
$Y_{ij}$, where \textit{i} represents an individual and \textit{j} the month
since licensure, using the logistic regression model shown in (\ref{Ymodel}):
%
\begin{eqnarray}
\label{Ymodel} Y_{ij} & \sim&\operatorname{Bernoulli}(\pi_{ij}),
\nonumber
\\[-8pt]
\\[-8pt]
\nonumber
\operatorname{logit}(\pi_{ij}) &=& \log(\mathrm{m}_{ij})+
\alpha_{0} +\alpha _{1}b_{ij}+
\alpha_2 u_i,
\end{eqnarray}
where the $\log(\mathrm{m}_{ij})$ is an offset to account for the miles
driven during a particular month. Treatment of the CNC outcome as
binary is not problematic since more than 98\% of the monthly
observations had one or fewer CNCs observed. Although the $\log$ link
is ideal for count data, it is a reasonable correction for the miles
driven for the $\operatorname{logit}$ link when the risk of a crash is low.
An alternative parameterization would be to include the miles driven as
a covariate in the model. The parameter $\alpha_1$ assesses the
odds-ratio of a crash or near crash event when in the \textit{poor} versus
\textit{good} driving state; this odds ratio is simply $e^{\alpha_1}$,
while $\alpha_2$ reflects unaccounted for covariates beyond the hidden
state. The $X_{ij}$ are the sum of the observed elevated g-force count
data and this model incorporates heterogeneity among subjects
introducing the random effect in the mean structure shown in (\ref
{Xmodelhidden}):
%
\begin{eqnarray}
\label{Xmodelhidden} X_{ij} & \sim& \operatorname{Poisson}(\mu_{ij}),
\nonumber
\\[-8pt]
\\[-8pt]
\nonumber
\log(\mu_{ij})&=& \log(\mathrm{m}_{ij}) +
\beta_{0} + \beta _{1}b_{ij} +
\beta_{2}t_{j} + \beta_3 u_{i},
\end{eqnarray}
where $u_{i}$ is the random effect with Gaussian distribution and $t_j$
reflects the month of observation since licensure for a particular
individual (note that these observations are equally spaced and $t_j$
was not statistically significant when included in the CNC process,
hence its omission in this part of the model). Here the parameter
$\beta
_3$ in (\ref{Xmodelhidden}), along with the variance of the random
effect distribution, accounts for any variation not explained by the
other terms included in the model and induces a correlation between
outcomes. Next, we assume that $\{b_{ij}|u_i\}_{j=1}^{n_i}$ is a Markov
chain and that $b_{ij}|u_i$ is independent of $b_{it}|u_i$ for $j \neq
t$. The transition probabilities for the Markov chain must lie between
0 and 1, and the sum of transitioning from one state to either state
must be 1. Hence, the transition probabilities are modeled as
%
\begin{eqnarray}
\label{p10r} &&\mathrm{p}_{01}(u_{i})\dvtx\quad \operatorname{logit}
\bigl( \operatorname {Pr}(b_{ij}=1|b_{i,j-1}=0|u_i)
\bigr)= \gamma_{01}+\delta _1y_{i,j-1}+u_{i},
\nonumber
\\[-8pt]
\\[-8pt]
\nonumber
&&\mathrm{p}_{10}(u_{i})\dvtx\quad \operatorname{logit} \bigl(
\operatorname {Pr}(b_{ij}=0|b_{i,j-1}=1|u_i) \bigr)=
\gamma_{10} + \delta_2 y_{i,j-1}+
\delta^{\ast} u_{i},
\end{eqnarray}
where the parameter $\delta^{\ast}$ in (\ref{p10r}) characterizes the
degree of correlation between transition probabilities among
individuals. Two different types of correlations are described by
$\delta$. If $\delta^{\ast} < 0$, then this implies individuals
have a
tendency to remain in either state. If $\delta^{\ast} > 0$, then this
implies some individuals exhibit a tendency to transition more between
states than others. This approach to describe the transitions in
similar to that presented by \citet{AlbertFollman}.
Introducing random effects in this manner is of great benefit
computationally, however, there is a downside risk in that this
implementation assumes the processes are highly correlated. This is
especially important for the hidden process where if the correlation
between transition probabilities is not very strong, biased estimates
will result. We present these findings in our simulation section.
\subsection{Estimation}
Letting ${\Psi}$ represent all parameters included in the model
discussed above, the likelihood for the joint model is
%
\begin{eqnarray}
\label{mhmmlike} &&L(\Psi; \mathbf{y},\mathbf{x})
\nonumber\\
&&\qquad=\int_{\mathbf{u}}\sum_{\mathbf{b}}f
\bigl(\mathbf{y}|(\mathbf{b},\mathbf{u}),\Psi \bigr)g\bigl(\mathbf{x}|(\mathbf{b}|
\mathbf{u}),\Psi\bigr)h(\mathbf{u};\Psi )\,d\mathbf {u}
\nonumber
\\[-8pt]
\\[-8pt]
\nonumber
&&\qquad=\int_{\mathbf{u}}\sum_{\mathbf{b}}
\Biggl\{ \prod_{i=1}^{N}\prod
_{j=2}^{n_{i}}f\bigl(y_{ij}|(b_{ij}|
\mathbf{u}),\Psi \bigr)g\bigl(x_{ij}|(b_{ij}|\mathbf {u}),\Psi
\bigr) \Biggr\}\\
&&\hspace*{59pt}{}\times \Biggl\{ \prod_{i=1}^{N}r_{i0}
\prod_{j=3}^{n_{i}}\mathrm {p}_{b_{i,j-1},b_{ij}|\mathbf{u}}
\Biggr\}h(\mathbf{u};\Psi)\,d \mathbf{u},\nonumber
\end{eqnarray}
where the summation associated with $\mathbf{b}$ represents all
possible state sequences for an individual and the initial state
probabilities are given by $\{r_{i0}\}$ and may include a
subject-specific random effect. In (\ref{mhmmlike}) we assume the
crash/near crash and kinematic event data are conditionally
independent. We also assume the $\{u_{i}\}$ are independent and
identically distributed and the observations for any driver are
independent given the random effect $u_{i}$ and the sequence of hidden
states. Given these assumptions, the likelihood given in (\ref
{mhmmlike}) simplifies to a product of one-dimensional integrals shown
in (\ref{finlike}):
%
\begin{eqnarray}
\label{finlike} &&L(\Psi; \mathbf{y},\mathbf{x})\nonumber
\\
&&\qquad=\prod_{i=1}^{N}\int
_{u_i} \Biggl\{\sum_{\mathbf
{b}_{i}}r_{i0}f
\bigl(y_{i2}|(b_{i2}|u_i),\Psi\bigr)g
\bigl(x_{i2}|(b_{i2}|u_i),\Psi\bigr)
\nonumber
\\[-8pt]
\\[-8pt]
\nonumber
&&\hspace*{33pt}\qquad\quad{}\times\prod_{j=3}^{n_{i}}\mathrm
{p}_{b_{i,j-1},b_{ij}|u_i}f\bigl(y_{ij}|(b_{ij}|u_i),
\Psi \bigr)\\
&&\hspace*{163pt}{}\times g\bigl(x_{ij}|(b_{ij}|u_i),\Psi\bigr)
\Biggr\}h(u_i;\Psi) \,du_{i}.\nonumber
\end{eqnarray}

The different roles the random effects perform in this joint model are
of particular interest. The purpose of the inclusion of the random
effect in the conditional model for the kinematic data $\{
({x}_i|(b_i|u_i))\}$ provides a relaxation of the assumption that the
observations for an individual are conditionally independent given the
hidden states $\{\mathbf{b}_i\}$ as well as accounting for
overdispersion for the kinematic event data. The inclusion of the
random effect in the transition probabilities allows the transition
probabilities to vary across individuals, inducing a correlation
between transition probabilities that induces a relationship between
the kinematic events and CNC processes. Further, the random effect
provides a departure from the assumption that the transition process
follows a Markov chain. One could formulate a reduced model that
includes a random effect only in the hidden process, in the hidden
process and one or both observed processes, or only in observed processes.

Several possibilities exist for maximizing the likelihood given by
(\ref
{finlike}). Two common approaches are the Monte Carlo expectation
maximization algorithm introduced by \citet{Wei} and the simulated
maximum likelihood methods discussed by \citet{Mccul}. In Section~\ref{sec2.3},
we propose a different method for parameter estimation that does not
rely on Monte Carlo methods which are difficult to implement and
monitor for convergence. Our method utilizes the \textit{forward--backward}
algorithm to evaluate the individual likelihoods conditional on the
random effect. As we will show, incorporation of this algorithm
provides a simpler means of computing the posterior probability of the
hidden state at any time point than MCEM. Further, this method provides
a straightforward approach for likelihood and variance evaluation. This
approach has the added benefit of producing the estimated
variance--covariance matrix for parameter estimates as determined by the
inverse of the observed information matrix.

\subsection{Maximizing the likelihood using an implementation of the
\textit{forward--backward} algorithm}\label{sec2.3}
In maximizing the likelihood given by (\ref{finlike}), our approach
first evaluates the portion of the likelihood described by the observed
data given the random effect and hidden states shown here:
%
\begin{eqnarray}
&&\Biggl\{\sum_{\mathbf{b}_{i}}r_{i0}f
\bigl(y_{i2}|(b_{i2}|u_i),\Psi \bigr)g
\bigl(x_{i2}|(b_{i2}|u_i),\Psi\bigr)
\nonumber
\\[-8pt]
\\[-8pt]
\nonumber
&&\qquad{}\times\prod
_{j=3}^{n_{i}}\mathrm {p}_{b_{i,j-1},b_{ij}|u_i}f
\bigl(y_{ij}|(b_{ij}|u_i),\Psi \bigr)g
\bigl(x_{ij}|(b_{ij}|u_i),\Psi\bigr) \Biggr\},
\end{eqnarray}
using the \textit{forward--backward} algorithm and subsequent numerical
integration over the random effect via adaptive Gaussian quadrature. We
then use a quasi-Newton method to maximize this result.
The alteration of the \textit{forward--backward} algorithm to accommodate
joint outcomes is described as follows. Here, let the vectors $\mathbf
{Y}_{ik}^{j}=(Y_{ik},\ldots,Y_{ij})'$ with realized values
$(y_{ik},\ldots
,y_{ij})'$ and $\mathbf{X}_{ik}^{j}=(X_{ik},\ldots,X_{ij})'$ with
realized values $(x_{ik},\ldots,x_{ij})'$. Decompose the joint
probability for an individual as follows:
%
\begin{eqnarray}
\label{third} &&\operatorname{Pr}(b_{ij}=m,\mathbf{Y}_{i}=
\mathbf{y}_{i},\mathbf {X}_{i}=\mathbf{x}_{i}|u_{i})\nonumber\\
&&\qquad=\operatorname{Pr} \bigl(Y_{i2}^{j}=y_{i2}^{j},X_{i2}^{j}=x_{i2}^{j},b_{ij}=m|u_{i}
\bigr)\\
&&\qquad\quad{}\times\operatorname {Pr}\bigl( Y_{i,j+1}^{n}=y_{i,j+1}^{n}|Y_{i2}^{j}=y_{i2}^{j},(b_{ij}=m|
u_{i})\bigr)
\nonumber
\\
&&\qquad\quad{} \times\operatorname{Pr}\bigl( X_{i,j+1}^{n}=x_{i,j+1}^{n}|X_{i2}^{j}=x_{i2}^{j},(b_{ij}=m|u_{i})
\bigr)\nonumber
\\
&&\qquad= \operatorname{Pr} \bigl(Y_{i2}^{j}=y_{i2}^{j},X_{i2}^{j}=x_{i2}^{j},b_{ij}=m|u_{i}
\bigr) \operatorname{Pr} \bigl(Y_{i,j+1}^{n}=y_{i,j+1}^{n}|(b_{ij}=m|
u_{i})\bigr)
\nonumber
\\
&&\qquad\quad{}\times\operatorname {Pr}\bigl(X_{i,j+1}^{n}=x_{i,j+1}^{n}|(b_{ij}=m|u_{i})
\bigr)
\nonumber
\\
&&\qquad= a_{im}(j)z_{im}(j)\qquad \mbox{for }m=0,1,\nonumber
\end{eqnarray}
where $a_{im}(j)$ and $z_{im}(j)$ are referred to as the \textit{forward}
and \textit{backward} quantities, respectively, and are
%
\begin{eqnarray}
a_{im}(j) &=&\operatorname {Pr}\bigl(Y_{i2}^{j}=y_{i2}^{j},X_{i2}^{j}=x_{i2}^{j},b_{ij}=m|u_{i}
\bigr)\qquad\mbox{for } j=2, \ldots, n_{i},\nonumber
\\
a_{im}(1) &=&\operatorname{Pr}\bigl((b_{i1}=m|u_i)
\bigr)\operatorname {Pr}\bigl(Y_{i2}=y_{i2}|(b_{i2}=m|u_{i})
\bigr)\nonumber\\
&&{}\times\operatorname{Pr}\bigl(X_{i2}=x_{i2}| (b_{i2}=m|u_{i})
\bigr),\nonumber
\\
z_{im}(j) &=&\operatorname {Pr}\bigl(Y_{i,j+1}^{n_{i}}=y_{i,j+1}^{n_{i}},X_{i,j+1}^{n_{i}}=x_{i,j+1}^{n_{i}},b_{ij}=m|u_{i}
\bigr)\nonumber \\
\eqntext{\mbox{for } j=1,\ldots,(n_{i}-1),}
\\
z_{im}(n_{i})&=& 1\qquad\mbox{for all } i.\nonumber
\end{eqnarray}
The $a_{im}(j)$ and $z_{im}(j)$ are computed recursively in $j$ by
using the following:
\begin{eqnarray*}
a_{im}(j) &=&\sum_{l=0}^{1}
\operatorname {Pr}\bigl(Y_{i2}^{j}=y_{i2}^{j},X_{i2}^{j}=x_{i2}^{j},b_{i,j-1}=l,b_{ij}=m,u_{i}
\bigr)
\\[-2pt]
\nonumber
&=&\sum_{l=0}^{1}\operatorname {Pr}
\bigl(Y_{i2}^{j-1}=y_{i2}^{j-1},X_{i2}^{j-1}=x_{i2}^{j-1},b_{i,j-1}=l|u_{i}
\bigr) \mathrm{p}_{lm|u_{i}}
\\[-2pt]
\nonumber
&&\hspace*{14pt}{}\times\operatorname {Pr}\bigl(Y_{ij}=y_{ij},X_{ij}=x_{ij}|(b_{ij}=m|u_{i})
\bigr)
\\[-2pt]
\nonumber
&=&\sum_{l=0}^{1}a_{il}(j-1)
\mathrm{p}_{lm|u_{i}}\operatorname{Pr}\bigl(Y_{ij}=y_{ij},X_{ij}=x_{ij}|(b_{ij}=m|u_{i})
\bigr)
\\[-2pt]
\nonumber
&=&\sum_{l=0}^{1}a_{il}(j-1)
\mathrm{p}_{lm|u_{i}}\operatorname{Pr}\bigl(Y_{ij}=y_{ij}|(b_{ij}=m|u_{i})
\bigr) \\[-2pt]
&&\hspace*{14pt}{}\times\operatorname{Pr}\bigl(X_{ij}=x_{ij}|(b_{ij}=m|u_{i})
\bigr)
\nonumber
\end{eqnarray*}
and
\begin{eqnarray*}
z_{im}(j) &=&\sum_{l=0}^{1}
\operatorname {Pr}\bigl(Y_{i,j+1}^{n_{i}}=y_{i,j+1}^{n_{i}},X_{i,j+1}^{n_{i}}=x_{i,j+1}^{n_{i}},b_{ij
+1}=l,(b_{ij}=m|u_{i})
\bigr)
\\[-2pt]
\nonumber
&=&\sum_{l=0}^{1}\operatorname {Pr}
\bigl(Y_{i,j+1}^{n_{i}}=y_{i1}^{n_{i}},X_{i,j+1}^{n_{i}}=x_{i,j+1}^{n_{i}}|(b_{ij
+1}=l|u_{i})
\bigr)\mathrm{p}_{ml|u_{i}}
\\[-2pt]
\nonumber
&&\hspace*{14pt}{}\times \operatorname{Pr} (Y_{i,j+1}=y_{i,j+1},X_{i,j+1}=x_{i,j+1}|b_{i,j+1}=l|u_{i})
\\[-2pt]
\nonumber
&=&\sum_{l=0}^{1}z_{il}(j+1)
\mathrm{p}_{ml|u_{i}}\operatorname {Pr}\bigl(Y_{i,j+1}=y_{i,j+1},X_{i,j+1}=x_{i,j+1}|(b_{i,j+1}=l|u_{i})
\bigr)
\\[-2pt]
\nonumber
&=&\sum_{l=0}^{1}z_{il}(j+1)
\mathrm{p}_{ml|u_{i}}\operatorname {Pr}\bigl(Y_{i,j+1}=y_{i,j+1}|(b_{i,j+1}=l|u_i)
\bigr)\\[-2pt]
&&\hspace*{14pt}{}\times\operatorname{Pr}\bigl(X_{i,j
+1}=x_{i,j+1}|(b_{i,j+1}=l|u_i)
\bigr).
\nonumber
\end{eqnarray*}
For any individual, the likelihood conditional on the random effect may
be expressed as a function of the forward probabilities, so for a
two-state Markov chain the conditional likelihood for an individual is
%
\begin{eqnarray}
L_{i|u_i} &=& \operatorname{Pr}(\mathbf{Y}_i=\mathbf{y}_i,
\mathbf {X}_i=\mathbf {x}_i|u_i)\nonumber
\\[-2pt]
&=& \sum_{l=0}^{1} \operatorname{Pr}(
\mathbf{Y}_i=\mathbf{y}_i,\mathbf{X}_i=
\mathbf {x}_i,b_{n_i}=l|u_i)
\\[-2pt]
\nonumber
&=&\sum_{l=0}^1a_{il}(n_i|
\Psi,u_i),
\end{eqnarray}
%
where $a_{i0}(n_i|\Psi,u_i) \mbox{ and } a_{i1}(n_i|\Psi,u_i)$ are the
\textit{forward} probabilities for subject $i$ associated with states 0
and 1, respectively, evaluated at the last observation of the subject's
observation sequence $n_i$.
The marginal likelihood for an individual can now be found by
integrating with respect to the random effect\looseness=-1
%
\begin{equation}
\label{iLikemhmm} L_i = \int_{u_i} \bigl
\{a_{i0}(n_i|\Psi,u_i) + a_{i1}(n_i|
\Psi ,u_i) \bigr\}h(u_i)\,du_i
\end{equation}
and the complete likelihood can be expressed as a product of individual
likelihoods.

\subsection{Numerical integration}
Adaptive Gaussian quadrature can be used to integrate (\ref
{iLikemhmm}). This technique is essential to obtaining accurate
parameter estimates, as the integrand is sharply ``peaked'' at different
values depending on the observed measurements of the individual.
Applying the results described in \citet{Gaussquad} to the joint hidden
Markov model, numerical integration of (\ref{iLikemhmm}) is achieved by
considering the distribution of the random effects to be $N(0,\theta
^2)$. The procedure for obtaining maximum likelihood estimates for
model parameters are shown in Table~\ref{MaxLikeProc}.

\begin{table}[b]
\caption{Procedure for obtaining maximum likelihood estimates for the
joint mixed hidden Markov model}
\label{MaxLikeProc}
\begin{tabular*}{\textwidth}{@{\extracolsep{\fill}}lp{330pt}@{}}
\hline
(1)& Select initial parameter estimates $p^0$. \\
(2)& Compute the set of adaptive quadrature points for each individual
$q_i$ given the current parameter estimates $p^{m}$.\\
(3)& Maximize the likelihood obtained using the \textit{forward--backward}
algorithm and adaptive quadrature via $q_i \in Q$ using the quasi-Newton method.\\
(4)& Update parameter estimates $p^{(m+1)}$.\\
(5)& Repeats steps (2)--(4) until parameters converge.\\
\hline
\end{tabular*}
\end{table}

\subsection{Estimation of posterior hidden state probabilities}
The \textit{forward--\break backward} implementation in evaluating the likelihood
is not only efficient (the number of operations to compute the
likelihood conditional on the random effect is of linear order as the
observation sequence increases), but it also provides a mechanism for
recovering information about the hidden states. By leveraging the
\textit{forward} and \textit{backward} probabilities, we can compute the hidden
posterior state probabilities $\{\hat{b}_{ij}\}$:
%
\begin{eqnarray}
\label{bgivY} \mathrm{E}(b_{ij}|{y_i,x_i})&=&
\mathrm{E}_{u_i|{y}_i,x_i}\bigl\{\mathrm {E}(b_{ij}|{y_i,x_i},u_i)
\bigr\}\nonumber
\\
&=&\int_{u_i} \bigl\{\operatorname{Pr}
\bigl(b_{ij}=1|({y_i,x_i},u_i)
\bigr) \bigr\} h_{u_i|({y_i,x_i})}\,du_i
\\
\nonumber
&=&\int_{u_i}
\biggl\{\frac{\operatorname{Pr}(b_{ij}=1,(\mathbf
{y}_i,\mathbf
{x}_i),u_i)}{\operatorname{Pr}(\mathbf{y}_i,\mathbf{x}_i)} \biggr\}\,du_i,
\end{eqnarray}
since
\begin{eqnarray*}
\operatorname{Pr}\bigl(b_{ij}=1,(\mathbf{y}_i,\mathbf{x}_i),u_i
\bigr)&=&\operatorname {Pr}\bigl(b_{ij}=1,(\mathbf{y}_i,\mathbf{x}_i)|u_i
\bigr)h(u_i)
\\
&=&a_{i1}(j)z_{i1}(j)h(u_i),
\end{eqnarray*}
equation (\ref{bgivY}) can be expressed as
%
\begin{eqnarray}
\label{bgivYfin}&& \int_{u_i} \biggl\{\frac{\operatorname{Pr}
(b_{ij}=1,(\mathbf{y}_i,\mathbf
{x}_i),u_i)}{\operatorname{Pr}(\mathbf{y}_i,\mathbf{x}_i)} \biggr
\}\,du_i
\nonumber
\\[-8pt]
\\[-8pt]
\nonumber
&&\qquad=\int_{u_i} \biggl\{\frac{a_{i1}(j)z_{i1}(j)h(u_i)}{ \{\int_{u_i}(a_{i0}(j)+a_{i1}(j))h(u_i)\,du_i \}}
\biggr\}\,du_i.
\end{eqnarray}
Evaluation of (\ref{bgivYfin}) is accomplished via adaptive Gaussian
quadrature as outlined in the earlier section and the quantities of
interest in (\ref{bgivYfin}) are readily available after running the
\textit{forward--backward} algorithm.

Using a shared random effect is also attractive in that it is
computationally more efficient than incorporating separate random
effects for the count outcome and transition probabilities. Generally
speaking, once the number of random effects exceeds three or four
(depending on the type of quadrature being used and the number of nodes
included for each integration), direct maximization is no longer a
computationally efficient method and Monte Carlo expectation
maximization (MCEM) is an appealing alternative approach. In accounting
for heterogeneity with a single random effect, we eliminate the need
for MCEM.

\section{Simulation of the mixed model}
\label{sec3}
We performed a simulation to investigate the performance of parameter
estimation using the proposed approach. Under the shared random effect
parameterization, we use the model in (\ref{simhmmre}) for the simulation
%
\begin{eqnarray}
\label{simhmmre} \operatorname{logit}(\pi_{ij}) &=& \alpha_{0} +
\alpha_{1}b_{ij} +\alpha_{2} u_i,
\nonumber
\\
\log(\mu_{ij}) &=& \beta_{0}+\beta_{1}b_{ij}
+ \beta_2 u_i,
\nonumber
\\
\operatorname{logit}\bigl\{ \operatorname{Pr}(b_{ij}=1|b_{ij-1}=0) \bigr\}
&=&\gamma_{01} + \delta_1 y_{i,j-1} +
u_i,
\\
\operatorname{logit}\bigl\{ \operatorname{Pr}(b_{ij}=1|b_{ij-1}=0) \bigr\}
&=&\gamma_{10}+ \delta_2 y_{i,j-1} +
\delta^{\ast} u_{i},
\nonumber
\\
\operatorname{logit}(b_{i1}=1) &=& \pi_{1},
\nonumber
\end{eqnarray}
where $u_{i} \backsim N(0,e^{\lambda})$.

\begin{table}
\caption{Parameter estimates for mixed hidden Markov model 1000
simulations (60 individuals, 20 observations) using $Q=5$ and $Q=11$
quadrature points}
\label{simretable}
\begin{tabular*}{\textwidth}{@{\extracolsep{\fill}}ld{2.2}d{2.3}ccd{2.3}cc@{}}
\hline
& &\multicolumn{3}{c} {$\bolds{Q=5}$}&
\multicolumn{3}{c@{}}{$\bolds{Q=11}$}\\[-4pt]
& &\multicolumn{3}{c} {\hrulefill}&
\multicolumn{3}{c@{}}{\hrulefill}\\
\textbf{Parameter} & \multicolumn{1}{c}{$\bolds{\theta}$} & \multicolumn{1}{c}{$\bolds{\overline{\hat{\theta}}}$}&
\multicolumn{1}{c}{$\bolds{\hat{\theta
}_{\mathrm{sd}}}$}&\multicolumn{1}{c}{$\bolds{\sigma_{\hat{\theta}}}$} & \multicolumn{1}{c}{$\bolds{\overline{\hat{\theta}}}$} &
\multicolumn{1}{c}{$\bolds{\hat
{\theta}_{\mathrm{sd}}}$} &\multicolumn{1}{c@{}}{$\bolds{\sigma_{\hat{\theta}}}$}\\
\hline
$\alpha_{0}$ & -1.0 & -1.01 & 0.12& 0.12 & -1.01 & 0.11& 0.11 \\
$\alpha_{1}$ & 2.0 & 2.03 & 0.18& 0.19 & 2.03 & 0.18& 0.17 \\
$\alpha_{2}$ & 1.5 & 1.52 & 0.41& 0.42 & 1.51 & 0.42& 0.44 \\
$\beta_{0}$ & -1.0 &-1.01 & 0.10& 0.10 & -1.00 & 0.09& 0.06 \\
$\beta_{1}$ & 2.0 & 2.01 & 0.12& 0.09 & 2.02 & 0.11& 0.09 \\
$\beta_{2}$ & 0.25 & 0.255 & 0.06& 0.07 & 0.251 & 0.06& 0.04 \\
$\gamma_{01}$ & -0.62 & -0.61 & 0.19& 0.18 & -0.62 & 0.19& 0.17 \\
$\gamma_{10}$ & 0.4 & 0.42 & 0.34& 0.32 & 0.42 & 0.30& 0.30 \\
$ \lambda$ & 0.0 & -0.03 & 0.16& 0.15 & -0.03 & 0.16& 0.16 \\
$\delta^{\ast}$ & 2.00 & 2.15 & 0.45& 0.41 & 2.02 & 0.43 & 0.42 \\
$\delta_1$ & 1.00 & 1.08 & 0.22& 0.21 & 1.02 & 0.20 & 0.21 \\
$\delta_2$ & 3.00 & 2.97 & 0.41& 0.43 & 2.97 & 0.44 & 0.42 \\
$\pi_{i1}$ & -0.8 & -0.82 & 0.06& 0.05 & -0.81 & 0.05 & 0.05 \\
\hline
\end{tabular*}
\end{table}

The simulations were conducted with 20 observations on 60 subjects.
Using a 1.86-GHz Intel Core 2 Duo processor, the fitting of the shared
model took less than 3 minutes on average. The simulation results (1000
simulations) are shown in Table~\ref{simretable}. Parameter estimates
obtained using adaptive Gaussian quadrature with five and eleven
points, respectively, are presented along with true parameter value
$(\theta)$ and mean $(\overline{\hat{\theta}})$, the sample standard
deviation for the parameter estimates $\hat{\theta}_{\mathrm{sd}}$, and the
average asymptotic standard errors $\sigma_{\hat{\theta}}$. In
performing the estimation using five quadrature points, parameter
estimation was quite accurate with the exception of the coefficient of
the random effect ${\delta^{\ast}}$ in the 0--1 transition with an
average estimated value of 2.15 compared to the actual value of 2.0.
Other parameters display very little bias. The bias for ${\delta^\ast}$
virtually disappears when performing the integration via adaptive
Gaussian quadrature using ten points where the average estimated value
was ${\delta^\ast}=2.02$. These results were unchanged when evaluating
for possible effects due to the total number of subjects or
observations (varying $n$ and~$I$). Additionally, the average
asymptotic standard errors agree quite closely to the sample standard
deviations for all model parameters. Similar results hold for the case
where random effects are only included in the hidden process. We used
different starting values to examine the sensitivity of estimation to
initial values, and our proposed algorithm was insensitive to the
selection of these values. Simulation results provide support that the
complexity of the model does not inhibit parameter estimation,
rendering discovery of heterogeneity in the observed and hidden
processes as a nice byproduct for the model.

\begin{table}[t]
\caption{Parameter estimates for the hidden process when the true
underlying random effects distribution is correlated. 1000 simulations
(60 individuals, 20 observations)}
\label{SIMcorr}
\begin{tabular*}{\textwidth}{@{\extracolsep{\fill}}ld{2.3}d{2.3}d{2.3}d{2.3}d{2.2}d{2.2}d{2.2}@{}}
\hline
\multicolumn{1}{@{}l}{\textbf{Parameter}} & \multicolumn{1}{c}{\textbf{True value}}& \multicolumn{1}{c}{$\bolds{\rho=1}$} &
\multicolumn{1}{c}{$\bolds{\rho=0.95}$}& \multicolumn{1}{c}{$\bolds{\rho=0.9}$}& \multicolumn{1}{c}{$\bolds{\rho
=0.8}$}& \multicolumn{1}{c}{$\bolds{\rho=0.7}$}& \multicolumn{1}{c@{}}{$\bolds{\rho=0.6}$}\\
\hline
$\gamma_{10}$ & -0.62 & -0.620 & -0.628 & -0.640 & -0.67 & -0.69 &
-0.72 \\
$\gamma_{01}$ & 0.4 & 0.401 & -0.399 & 0.398 & 0.32 & 0.28 & 0.26\\
$\delta$ & 2.0 & 2.00 & 2.11 & 2.28 & 2.44 & 2.61 & 2.68\\
\hline
\end{tabular*}
\end{table}

While performance of the estimation procedure for the mixed hidden
Markov model in (\ref{simhmmre}) was quite good, this model formulation
assumes near perfect correlation between the random effects. To examine
the robustness of the estimation procedure to this assumption, we
evaluated model performance in the case where there are correlated
random effects in the hidden process. For this case, data was simulated
for the transition probabilities using (\ref{corrtrans}):
%
\begin{eqnarray}
\label{corrtrans} \operatorname{logit}\bigl\{ \operatorname{Pr}(b_{ij}=1|b_{ij-1}=0)
\bigr\}&=&\gamma_{01} + u_1,
\nonumber
\\[-8pt]
\\[-8pt]
\nonumber
\operatorname{logit}\bigl\{ \operatorname{Pr}(b_{ij}=1|b_{ij-1}=0) \bigr
\}&=&\gamma_{10}+ u_{2},
\end{eqnarray}
where $(u_1,u_2)$ follow a bivariate normal distribution $\mathit{BVN}(\mathbf
{0},\bolds{\Sigma})$. Our model was then fit to the simulated data
using the parameterization in (\ref{mfit}):
%
\begin{eqnarray}
\label{mfit} \operatorname{logit}\bigl\{ \operatorname{Pr}(b_{ij}=1|b_{ij-1}=0)
\bigr\}&=&\gamma_{01} + u_i,
\nonumber
\\[-8pt]
\\[-8pt]
\nonumber
\operatorname{logit}\bigl\{ \operatorname{Pr}(b_{ij}=1|b_{ij-1}=0) \bigr
\}&=&\gamma_{10} + \delta u_{i}.
\end{eqnarray}
The results for this simulation are shown in Table~\ref{SIMcorr}. For
moderate departures from perfect correlation, biased estimates result
in the hidden process. Thus, we recommend first considering correlated
random effects in the hidden process before use of the shared random
effects model. In the case of our application, the transition process
exhibited very high correlation, so we proceed in the analysis with our
estimation procedure. Details for estimation using bivariate adaptive
Gaussian quadrature are shown in the supplementary material [\citet{jaz}].

\section{Results}
\label{sec4}
The two-state mixed hidden Markov model presented in the previous
sections was applied to the NTDS data. Table~\ref{mHmmresults} displays
the parameter estimates and associated standard errors. The initial
probability distribution for the hidden states was common to all
individuals in the study and modeled using $\operatorname{logit}(r_1)=\pi_1$
and the random effects distribution was $\mathrm{N}(0,e^\lambda)$.
Several models were compared based on the relative goodness-of-fit
measure, AIC. Model excursions included evaluating the suitability of a
three-state hidden Markov model without random effects $(\mathit{AIC}=3563.06)$
(model shown in the supplementary material [\citet{jaz}]), the two-state model without random
effects $(\mathit{AIC}=3548.97)$, and the two-state model with random effects
$(\mathit{AIC}=3492.73)$.

\begin{table}
\tablewidth=250pt
\caption{Parameter estimates for the mixed hidden Markov model as
applied to the NTDS data}
\label{mHmmresults}
\begin{tabular*}{250pt}{@{\extracolsep{\fill}}ld{2.3}d{1.3}@{}}
\hline
\multicolumn{1}{@{}l}{\textbf{Parameters}} & \multicolumn{1}{c}{\textbf{Estimate}} & \multicolumn{1}{c@{}}{\textbf{Std Err}}  \\
\hline
$\alpha_{0} $ & -7.48 & 0.14  \\
$\alpha_{1} $ & 1.49 & 0.25  \\
$\alpha_2 $ & 0.03 & 0.02  \\
$\beta_{0} $ & -5.97 & 0.13  \\
$\beta_{1} $ & 1.31 & 0.06  \\
$\beta_{2} $ &0.007 &0.004  \\
$\beta_{3}$ & 1.10 & 0.04  \\
$\lambda$ &-0.18 & 0.33  \\
$\delta^\ast$ & 1.25 & 0.32  \\
$\gamma_{10} $ & -2.13 & 0.35 \\
$\gamma_{01} $ & -3.47 & 0.28  \\
$\delta_1$ & 1.75 & 0.24  \\
$\delta_2$ & -2.17 & 0.53  \\
$\pi_1 $ &-0.83 &0.28  \\
\hline
\end{tabular*}
\end{table}

The fixed effects model provided initial parameter estimates for most
model parameters while multiple starting values for $\lambda$ were used
in conjunction with a grid search over parameters $\beta_3$ and
$\delta
^\ast$ to determine these initial values. The number of quadrature
points implemented at each iteration was increased until the likelihood
showed no substantial change. As illustrated in the simulation study,
there were eleven points used in the adaptive quadrature routine.
Standard error estimates were obtained using a numerical approximation
to the Hessian using the \textsc{\{nlm\}} package in R. The
coefficients of the hidden states $\alpha_1$ and $\beta_1$ are both
significantly greater than zero, indicating that drivers operating in a
\textit{poor} driving state $(b_{ij}=1)$ are more likely to have a
crash/near crash event and, correspondingly, a highly number of
kinematic events. While the variance component of the random effect is
somewhat small, the dispersion parameter $\beta_3$ is highly
significant, indicating the data are overdispersed. Interestingly,
heterogeneity is not exhibited in the CNC outcome as the coefficient
for the random effect $\alpha_2$ is insignificant, providing support to
the notion that the hidden state is capturing unobserved quantities in
a meaningful way. There is evidence of heterogeneity across individuals
in their propensity to change between states as indicated by $\lambda$
and $\delta^\ast$. In the case of the NTDS data, $\delta^\ast> 0$
indicates a positive correlation between the transition probabilities,
meaning that some individuals are prone to changing more often between
states than others. Coefficients in the hidden process, $\delta_1$ and
$\delta_2$, illustrate that transition between states depends on
previous CNC outcomes. A prior crash was associated with an increased
probability of transitioning from the \textit{good} driving state to the
\textit{poor} one ($\delta_1=1.75$) and a decreased probability of
transitioning from the \textit{poor} to the \textit{good} driving state
($\delta_2=2.17$). Since the shared random effect, which assumes a
perfect correlation between the random components, may not be robust to
a more flexible random effects structure, we also fit the model using
correlated random effects in the hidden process (see simulation
results). For a variety of starting values, the correlation coefficient
estimates were near 1 (0.998 or greater), giving us confidence in using
the shared random effect approach.

\begin{figure}[b]

\includegraphics{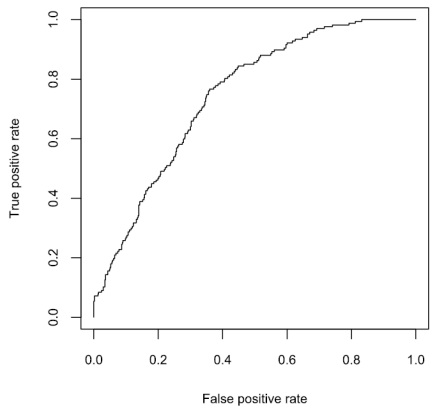}

\caption{ROC curve for the mixed hidden Markov model based one
``one-step ahead'' predictions (area under the curve ${}= 0.74$).}
\label{ROConestepmhmm}
\end{figure}

An interpretation of parameter estimates given in Table~\ref{mHmmresults} is
subject-specific and depends on a driver's exposure for a given month.
If we consider a subject driving the average mileage for all subjects
(358.1 miles), parameter estimates indicate that the risk of a
crash/near crash outcome increases from 0.16 to 0.47 when in the
\textit{poor} driving state, $b_{ij}=1$. Correspondingly, this ``average''
subject would also expect to experience 2.43 more kinematic events on
average when in the \textit{poor} driving state. For the typical teenager,
the likelihood of moving out of the \textit{poor} driving state decreases
from $10.6\%$ to $1.3\%$ when experiencing a CNC event in the previous
month. Similarly, the likelihood of moving out of the \textit{good}
driving state increases from $3.01\%$ to $15.2\%$ when experiencing a
CNC event in the previous month. 

\begin{figure}[t]

\includegraphics{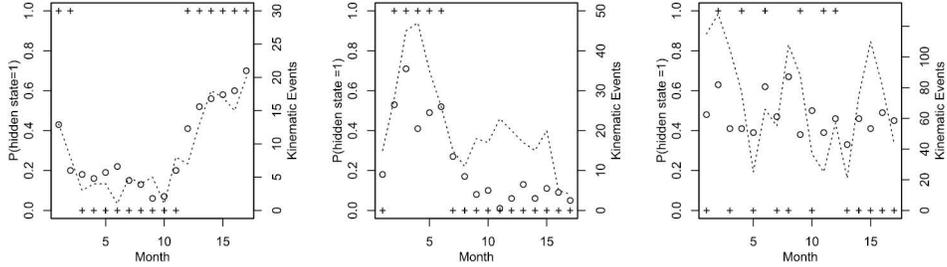}

\caption{Predicted value of the hidden state given the observed data
for three drivers. The ($\circ$) indicates the probability of being in
state 1 (poor driving), ($+$) indicates a crash/near crash event and
the dotted line indicates the composite kinematic measure.}\label{predbij}
\end{figure}

\begin{figure}[b]

\includegraphics{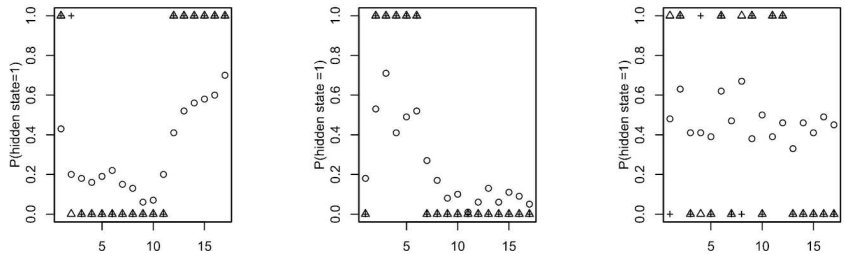}

\caption{Comparison of local and global decoding of the hidden states
and CNC outcomes. The ($\circ$) indicates the probability of being in
state 1 (poor driving), ($+$) indicates a CNC event and the
($\bigtriangleup$) indicates the hidden state occupation using the
Viterbi algorithm.}
\label{global}
\end{figure}

A receiver operating characteristic (ROC) curve was constructed to
determine the predictive capability of our model by plotting the true
positive rate versus the false positive rate for different cutoff
values. The ROC curve based on the ``one-step ahead'' predictions
[observed outcome given all previous kinematic observations $\operatorname
{Pr}(Y_{ij}=1|y_{i1},\ldots,y_{i,j-1},x_{i1},\ldots,x_{i,j-1})$] is
shown in Figure~\ref{ROConestepmhmm}. An attractive feature of our
model is that it allows for the development of a predictor based on
prior kinematic events. We constructed this ROC curve using a
cross-validation approach whereby one driver was removed from the data
set, model parameters were then determined using the remaining data and
these results were then used to predict the removed driver's crash/near
crash outcomes. The predictive accuracy of this model was moderately
high with an area under the curve of 0.74. Although the goodness of fit
was best for the two-state mixed hidden Markov model, area under the
curve for the other models was nearly identical.

A sample of three drivers and their corresponding hidden state
probability $\mathrm{E}_{u_i|\mathbf{y}_i,\mathbf{x}_i}\{\mathrm
{E}(b_{ij}|\mathbf{y}_i,\mathbf{x}_i,u_i)\}$ along with their
crash/near crash outcomes and total number of kinematic events is shown
in Figure~\ref{predbij}. It is evident how the total kinematic measures
influence the predicted value of the hidden state and work particularly
well for cases where driving is ``consistent'' over relatively short
time periods (i.e., low variation in kinematic measures for a given
time period). In cases where the driving kinematics exhibit a great
deal of variability, the model does not perform as well in predicting
crash/near crash outcomes as indicated by the rightmost panel in Figure~\ref{predbij}.
As a comparison, we show the results of global decoding
of the most likely hidden state sequence using the Viterbi algorithm in
Figure~\ref{global} for the same three drivers. Hidden state
classification is similar whether using global or local decoding for
the left two panels in Figure~\ref{global}, which is indicative of most
drivers in the study, while there are differences in the case of the
rightmost panel likely due to the greater variability in these data.

\section{Discussion}
\label{sec5}
In this paper we presented a mixed hidden Markov model for joint
longitudinal binary and count outcomes introducing a shared random
effect in the conditional model for the count outcomes and the model
for the hidden process. An estimation procedure incorporating the
\textit{forward--backward} algorithm with adaptive Gaussian quadrature for
numerical integration is used for parameter estimation. A welcome
by-product of the \textit{forward--backward} algorithm is the hidden state
probabilities for an individual during any time period. The shared
random effect eliminates the need for more costly numerical methods in
approximating the likelihood, such as higher dimensional Gaussian
quadrature or through Monte Carlo EM.

The model was applied to the NTDS data and proved to be a good
predictor of crash and near crash outcomes. Our estimation procedure
also provides a means of quantifying teenage driving risk. Using the
hidden state probabilities which represent the probability of being in
a \textit{poor} driving state given the observed crash/near crash and
kinematic outcomes, we can analyze the data in a richer way than
standard summary statistics. Additionally, our approach allows for a
broader class of predictors whereby the investigator may make
predictions based on observations that go as far into the past as warranted.

There are limitations to our approach. The shared random effect
proposes a rather strong modeling assumption in order to take advantage
of an appealing reduction in computational complexity. Using more
general correlated random effects approaches is an alternative, but
others have found that identification of the correlation parameter is
difficult [\citet{SmithMoffatt} and \citet{Alfo}]. Formal testing for
heterogeneity in these models is also a challenging problem [\citet{Altman2}].

There is also a potential issue of having treated the miles driven
during a particular month ($\mathrm{m}_{ij}$) as exogenous. For some
crashes, it is possible that previous CNC outcomes ($y_{i,j-1}$) may
affect the miles driven in the following month and our model does not
capture this dynamic. As with any study, greater clarity in the
information obtained for each trip might yield more valuable insights.
Metrics such as the type of road, road conditions and trip purpose,
while potentially useful, were not available for this analysis.

There are extensions to the model that may be useful. The model can
address more than two outcomes. We summarize the kinematic events at a
given time as the sum across multiple types. This approach could be
extended to incorporate multiple correlated processes corresponding to
each kinematic type. Depending on the situation, the additional
flexibility and potential benefits of such an extension may be worth
the increased computational cost.



 \section*{Acknowledgments}
We thank the Center for Information Technology, National Institutes of
Health, for providing access to the high-performance computational
capabilities of the Biowulf cluster computing system.
We also thank Bruce Simons-Morton for discussions related to this work.
Inquiries about the study data may be sent to P.~S. Albert at \printead*{e2}.

\begin{supplement}[id=suppA]
\stitle{Adaptive quadrature for the three-state mixed hidden Markov model}
\slink[doi]{10.1214/14-AOAS765SUPP} 
\sdatatype{.pdf}
\sfilename{aoas765\_supp.pdf}
\sdescription{We provide details on the adaptive quadrature routine for the MHMM with
bivariate normal random effects in the hidden process, as well as
expressions for the three-state hidden Markov model.}
\end{supplement}


\printaddresses
\end{document}